\documentclass[preprint2]{aastex}

\shorttitle{Zero Time Endian Conversion Technique}
\shortauthors{S. Eguchi}

\begin{document}

\title{``Superluminal'' FITS File Processing on Multiprocessors: Zero Time Endian Conversion Technique}

\author{Satoshi Eguchi\altaffilmark{1}}
\affil{Astronomy Data Center, National Astronomical Observatory of Japan, 2-21-1, Osawa, Mitaka, Tokyo 181-8588, Japan}
\email{satoshi.eguchi@nao.ac.jp}
\altaffiltext{1}{Postdoctoral Fellow of Japanese Virtual Observatory Project.}

\begin{abstract}
The FITS is the standard file format in astronomy, and it has been
extended to agree with astronomical needs of the day.
However, astronomical datasets have been inflating year by year.
In case of ALMA telescope, a $\sim$ TB scale 4-dimensional data cube
may be produced for one target.
Considering that typical Internet bandwidth is a few 10 MB/s at most,
the original data cubes in FITS format are hosted on a VO server,
and the region which a user is interested in should be cut out and
transferred to the user \citep{Eguchi2012}.
The system will equip a very high-speed disk array to process a TB scale
data cube in a few 10 seconds, and disk I/O speed, endian conversion and
data processing one will be comparable.
Hence to reduce the endian conversion time is one of issues to realize our
system.
In this paper, I introduce a technique named ``just-in-time endian conversion'',
which delays the endian conversion for each pixel just before it is
really needed, to sweep out the endian conversion time;
by applying this method, the FITS processing speed increases 20\% for single
threading, and 40\% for multi-threading compared to CFITSIO.
The speed-up by the method tightly relates to modern CPU architecture
to improve the efficiency of instruction pipelines due to break of
``causality'', a programmed instruction code sequence.
\end{abstract}


\section{Introduction}

The Flexible Image Transport System (FITS) is the standard data format for
astronomical observed data even though they are products through calibration
pipelines or otherwise.
One FITS file can store multiple CCD images and photon event lists as tables,
and this feature makes FITS format prevail from the radio band to
the X-ray band.
Especially, most archival datasets and source catalogs are provided
as FITS files in these days.

The original purpose of the FITS format was to transport digital astronomical
images from a computer to another with a magnetic tape \citep{Wells1981}.
There were no unified standard for computers at that time, and bit size
assigned to a character and an integer was quite different from one model
to another, even from the same makers.
Thus the authors newly had to create a machine independent and future expandable
image format for data exchange, FITS.
Since then the FITS format has been repeatedly extended to agree with astronomical
needs of the day \citep[e.g.,][]{Greisen1981,Grosbol1988}.

However, we will look at the issue of astronomical data inflation, not of the
format, in the years ahead;
Atacama Large Millimeter/submillimeter Array (ALMA), which is the largest radio
telescope built on the Chajnantor plateau in northern Chile, started observations
last year.
ALMA is estimated to generate $\sim$200\ TB observational raw data every year,
and the volume of a processed 4-dimensional data cube\footnote{$=$ (2D Image) $\otimes$ (Spectrum) $\otimes$ (Polarization)}
for one target may exceed $\gtrsim$2\ TB \citep{Lucas2004}.
Furthermore, Large Synoptic Survey Telescope (LSST), a project in 2020s, will
generate 30\ TB data every night\footnote{\url{http://www.lsst.org/lsst/science/development}}.
We need a system which assists astronomers to find something interested in such
big data.

Looking at such future, National Astronomical Observatory of Japan has
been developing a large data providing system for ALMA utilizing
the technology of Virtual Observatory (VO) to share our outputs
with global astronomical communities;
all processed datasets (FITS files) are hosted on a VO server, and
an user can select cut-out region to download by a web-based graphical user
interface \citep[Paper I hereafter]{Eguchi2012}.

A prototype service is already public\footnote{\url{http://jvo.nao.ac.jp/portal/alma/}},
and I am working on its optimization now.
The system has to process a TB scale data cube in a few 10 seconds for
users' convenience, thus it is planned to equip a very high-speed disk
array\footnote{A system which consists of 16 striping
solid state disks (SSDs) in the consumer products market effectively reaches
$\simeq$ 4\ GB/s read/write performance.} and disk I/O speed
and data processing one will be comparable.
All the components of the system consist of Intel platform, which adopts little
endian, while the FITS format does big endian.
For the interactive TB size FITS file processing system, the endian conversion time
is not negligible.

In this paper, I introduce a technique to make the endian conversion time
apparently disappear, and to make the system much faster by multiprocessing.
I describe the hardware and software configuration for evaluation in Section~2,
and compare endian conversion algorithms and their performance in Section~3.
In Section 4, I examine the best timing for endian conversion, and discuss
the performance increase by the conversion timing in Section 5.
Through the paper, I repeated measurements 100 times for each item, and
adopted its sample standard deviation (a square root of unbiased variance)
as 1-$\sigma$ statistical error, ignoring any systematic ones.

\section{Configuration and Test Data \label{sec-configuration}}

\begin{deluxetable}{ccc}
	\tabletypesize{\scriptsize}
	\tablecaption{The hardware and software information used for the evaluations\label{tab-configuration}}
	\tablewidth{0pt}
	\tablehead{
		\colhead{} & \colhead{Machine A} & \colhead{Machine B}
	}
	\startdata
		CPU & Intel Core i7-2600 (3.4\ GHz) & AMD FX-8350 (4.0\ GHz) \\
		RAM & 8\ GB ($8.80 \pm 0.03$\ GB/s) & 16\ GB ($22.04 \pm 0.07$\ GB/s) \\
		Storage & SSD (Read: $506.5 \pm 0.7$\ MB/s) & HDD (Read: $143 \pm 2$\ MB/s) \\
		Operating System & \multicolumn{2}{c}{Ubuntu 12.04.1 (amd64)} \\
		C/C++ Compiler & \multicolumn{2}{c}{GNU Compiler Collection Version 4.6} \\
		FITS Library & \multicolumn{2}{c}{CFITSIO Version 3.310} \\
	\enddata
	\tablecomments{
		Intel Turbo Boost and Hyper-Threading Technologies (for Machine A),
		AMD Turbo CORE Technology (for Machine B) are disabled through the paper.
		Hence 4 and 8 physical processors are available for Machine A and B,
		respectively.
	}
\end{deluxetable}

\begin{figure}
	\begin{center}
		\includegraphics[keepaspectratio, width=1.0\hsize, clip]{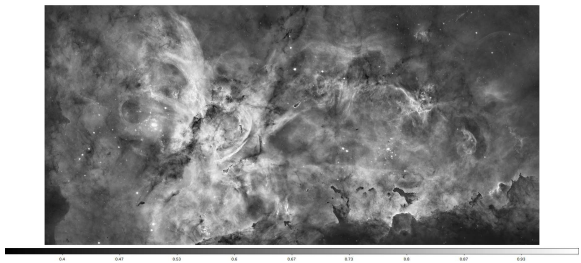}
	\end{center}
	\caption{Test FITS image to evaluate parallelization efficiencies:
	a false color mosaic image of Carina Nebula obtained with Hubble Space
	Telescope, consisting of 29,566$\times$14,321 pixels (3.4\ GB).
	\label{fig-test-image}}
\end{figure}

Table~\ref{tab-configuration} shows the hardware and software configuration
used for verification of the method.
I used two types of CPUs, Intel Core i7-2600 (for Machine A) and AMD FX-8350
(for Machine B), to prevent bias due to microarchitecture.
Through the paper, Intel Turbo Boost Technology (the former) and AMD Turbo
CORE Technology (the latter) are disabled by BIOS for simplicity.
In addition, Intel Hyper-Threading Technology (the former) is also disabled
for the same reason.
Thus Machine A and B are available 4 and 8 physical processors, respectively.
The memory bandwidths and storage speeds were obtained the following commands:
\texttt{dd if=/dev/zero of=/dev/null bs=1G count=100}, and
\texttt{hdparm -t (device)}, respectively.

The same software is installed in both computers:
Ubuntu 12.04.1 LTS (amd64), a Debian based 64-bit Linux, for operating
system, GNU Compiler Collection (GCC) Version 4.6 for C/C++ compiler
(\texttt{gcc}/\texttt{g++}), and CFITSIO Version 3.310 for C language FITS
library \citep{Pence2010}.
I applied the \texttt{-O2 -pipe -Wall} compile options to CFITSIO and programs
used in the paper.
The Streaming SIMD\footnote{Single Instruction/Multiple Data} Extensions 2 (SSE2)
codes in CFITSIO was enabled since I built the library on a 64-bit Linux\footnote{There
is no way to make the \texttt{\_\_SSE2\_\_} macro undefined with 64-bit GCC,
which switches the codes for SSE2 or otherwise in CFITSIO.}, but the
SSSE3 option was disabled since the SSSE3 instruction set is
treated as an extension in the amd64 environment.

I use a false color mosaic image of Carina Nebula obtained with Hubble Space
Telescope\footnote{\url{http://hubblesite.org/newscenter/archive/releases/2007/16/image/a/}}
for test data.
The image is public in Tagged Image File Format (TIFF), thus I converted it
into a gray scale double precision FITS file with \texttt{convert} command
provided by ImageMagick\footnote{\url{http://www.imagemagick.org/script/index.php}}.
The size is 29,566 pixels in width and 14,321 pixels in height.
The file volume is 3.4\ GB (Figure~\ref{fig-test-image}).
Through the paper, I put this FITS file on a tmpfs \citep{Rohland2001}
mounted on \texttt{/run/shm}, to ensure that the file is always on memory for
fast access.
See Appendix~\ref{sec-tmpfs} for the difference between tmpfs and ramdisk.

\section{Endian Conversion Algorithms}

\begin{figure}
	\begin{center}
		\includegraphics[keepaspectratio, width=1.0\hsize, clip]{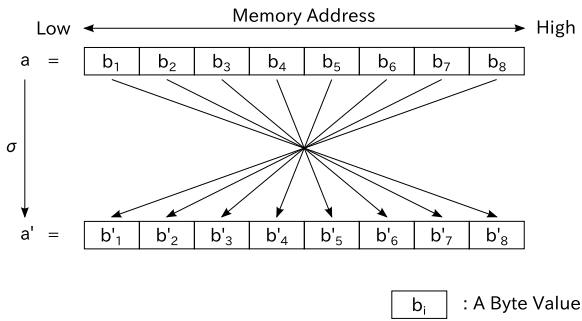}
	\end{center}
	\caption{The schematic diagram of a permutation operator $\sigma$ for the
	endian conversion of a 64-bit value.
	\label{fig-operator-sigma}}
\end{figure}

\subsection{Formalism}

Let $\left( b_{1}, b_{2}, \cdots ,b_{8} \right)$ be a byte sequence of an internal expression
of a 64-bit size value $a$.
The 64-bit endian conversion of $a$ can be expressed with a permutation
$\sigma$ as
\begin{equation}
a' = \left( b_{\sigma\left( 1 \right)}, b_{\sigma\left( 2 \right)}, \cdots ,b_{\sigma\left( 8 \right)} \right), \label{eq-endian-conversion1}
\end{equation}
where
\begin{equation}
\sigma = \left(
		\begin{array}{cccccccc}
		1 & 2 & 3 & 4 & 5 & 6 & 7 & 8 \\
		8 & 7 & 6 & 5 & 4 & 3 & 2 & 1
		\end{array}
		\right) \label{eq-endian-conversion2}
\end{equation} in Cauchy's two-line notation,
and $\sigma^{2} = {\bf 1}$ (Figure~\ref{fig-operator-sigma}).

\subsection{Implementation}

\subsubsection{Byte Shuffle: Straightforward Implementation}

A straightforward implementation of Eq.~(\ref{eq-endian-conversion1}) and
Eq.~(\ref{eq-endian-conversion2}) can be written as follows:
\begin{verbatim}
uint64_t byte_shuffle(uint64_t a)
{
  unsigned char *p = (unsigned char *)&a;
  unsigned char tmp;

  tmp = p[7]; p[7] = p[0]; p[0] = tmp;
  tmp = p[6]; p[6] = p[1]; p[1] = tmp;
  tmp = p[5]; p[5] = p[2]; p[2] = tmp;
  tmp = p[4]; p[4] = p[3]; p[3] = tmp;

  return a;
}
\end{verbatim}

I now call this method ``byte shuffle''.
One will find a short discussion about another implementation of byte shuffle
algorithm in Appendix~\ref{sec-another-byteshuffle}.

\subsubsection{Bit Shift \label{sec-definition-bit-shift}}

Another implementation to perform endian conversion is to use both bit shift
and logical operations:
\begin{verbatim}
uint64_t bit_shift(uint64_t a)
{
  return   ((a & 0x00000000000000FFULL)
             << 56)
         | ((a & 0x000000000000FF00ULL)
             << 40)
         | ((a & 0x0000000000FF0000ULL)
             << 24)
         | ((a & 0x00000000FF000000ULL)
             << 8)
         | ((a & 0x000000FF00000000ULL)
             >> 8)
         | ((a & 0x0000FF0000000000ULL)
             >> 24)
         | ((a & 0x00FF000000000000ULL)
             >> 40)
         | ((a & 0xFF00000000000000ULL)
             >> 56);
}
\end{verbatim}

Hereafter, I call this method ``bit shift''.

\subsubsection{\texttt{BSWAP}}

Intel i486 and later processors have the \texttt{BSWAP} instruction,
which converts the endian on a given 32-bit register.
The instruction is extended in order to accept a 64-bit register in amd64
\citep{Intel2012}.
Furthermore, GCC Version 4.3 and later have a helper function to call
the instruction, and its prototype is
\texttt{uint64\_t \_\_builtin\_bswap64(uint64\_t x);}.
Now I call endian conversions utilizing this function ``\texttt{BSWAP}''.

\subsubsection{SSE2}

SSE2 is a set of vector instructions for Intel platform,
became a part of default instruction set for amd64 environment.
The endian conversion codes utilizing SSE2 can process two 64-bit
values at once, and be written as follows:
\begin{verbatim}
#include <emmintrin.h>
void sse2(uint64_t a[2])
{
  __m128i r0 = _mm_load_si128((__m128i *)a);
                // r0 <- a
  __m128i r1 = _mm_srli_epi16(r0, 8);
                // 8-bit shifts towards right
                // for four 2-byte integers
  __m128i r2 = _mm_slli_epi16(r0, 8);
                // 8-bit shifts towards left
                // for four 2-byte integers
  r0 = _mm_or_si128(r1, r2);
                // 128-bit or operation
                // on r1 and r2
  r0 = _mm_shufflelo_epi16(r0,
               _MM_SHUFFLE(0, 1, 2, 3));
                // byte shuffle for the
                // lower half of r0 register
  r0 = _mm_shufflehi_epi16(r0,
               _MM_SHUFFLE(0, 1, 2, 3));
                // byte shuffle for the
                // higher half of r0 register
  _mm_store_si128((__m128i *)a, r0);
                // a <- r0
}
\end{verbatim}

There are almost the same codes in CFITSIO and
SLLIB/SFITSIO\footnote{\url{http://www.ir.isas.jaxa.jp/~cyamauch/sli/index.html}}.
I call these codes simply ``SSE2'', hereafter.

\subsection{SSSE3}

Another vector instruction set called ``SSSE3'' is available for Intel Core series
and later CPUs.
Utilizing this instruction set, one can perform endian conversion of two 64-bit values
at one instruction.
An example is follows:
\begin{verbatim}
#include <tmmintrin.h>
void ssse3(uint64_t a[2])
{
  static const __m128i mask
    = _mm_set_epi8(
        8, 9, 10, 11, 12, 13, 14, 15,
        0, 1, 2 ,3, 4, 5, 6, 7
      );
  __m128i r = _mm_load_si128((__m128i *)a);
  __m128i r = _mm_shuffle_epi8(r, mask);
  _mm_store_si128((__m128i *)a, r);
}
\end{verbatim}

There are almost same codes in CFITSIO too.
I call these codes simply ``SSSE3'', hereafter.

\begin{deluxetable}{cccccc}
	\tabletypesize{\scriptsize}
	\tablecaption{Endian Conversion Time\label{tab-endian-conversion-algorithm}}
	\tablewidth{0pt}
	\tablehead{
		\colhead{Machine} & \colhead{Bit Shift (msec)} & \colhead{\texttt{BSWAP} (msec)} & \colhead{SSE2 (msec)} & \colhead{SSSE3 (msec)} & \colhead{Byte Shuffle (msec)}
	}
	\startdata
		Machine A & $410 \pm 2$ & $410 \pm 2$ & $405 \pm 2$ & $372.4 \pm 0.1$ & $3190.3 \pm 0.6$ \\
		Machine B & $601.3 \pm 0.6$ & $605.8 \pm 0.7$ & $582.3 \pm 0.7$ & $598 \pm 2$ & $8056.3 \pm 0.3$ \\
	\enddata
	\tablecomments{The endian conversion time of 423,414,686 ($=$29,566$\times$14,321)
	\texttt{double}-type elements with various algorithms.
	}
\end{deluxetable}

\subsection{Benchmark \label{sec-endian-benchmark}}

To see which one is fastest and how they behave towards parallelization,
I performed simple benchmark.
In the benchmark, I reserved a \texttt{double}-type array whose number
of elements were set to 29,566$\times$14,321 = 423,414,686, just the
number of pixels in Figure~\ref{fig-test-image}, and filled the array for
uniform real random numbers of 32-bit resolution on $[-1000, 1000]$
generated with Mersenne Twister \citep{Matsumoto1998}.

\subsubsection{Single Thread \label{sec-endian-single-thread}}

The results are summarized in Table~\ref{tab-endian-conversion-algorithm}.
For Machine A, three algorithms except for SSSE3 and byte shuffle process
the test data in about 410 milliseconds, while SSSE3 does about 370
millisecond.
On the other hand, for Machine B, four algorithms except for byte shuffle
process the test data in about 600 milliseconds, and the SSE2 algorithm is
fastest in the all ones.
The byte shuffle algorithm is slowest by one oder compared to the others.

\begin{figure*}
	\begin{center}
		\includegraphics[keepaspectratio, width=0.75\hsize, clip]{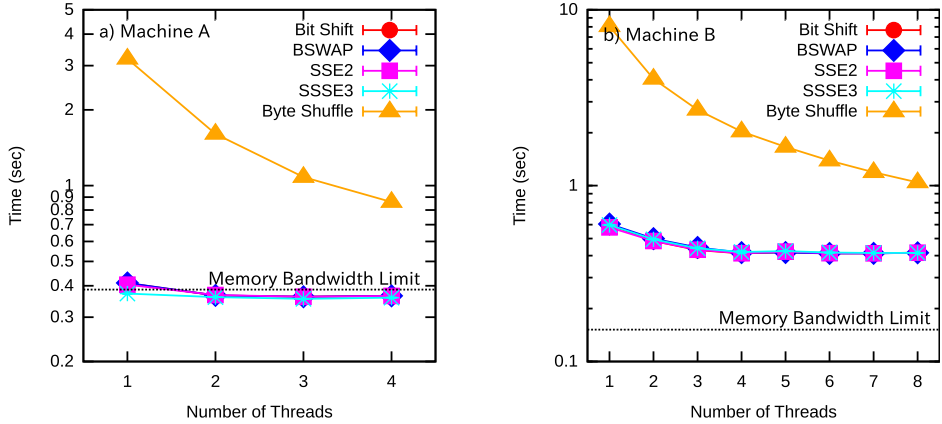}
	\end{center}
	\caption{CPU-scalability comparison between the endian coversion algorithms.
	All the algorithms except for byte shuffle seem to behave in the same way
	and scale up to only the half number of CPU cores due to the hardware
	I/O limits.
	\label{fig-endian-algorithm}}
\end{figure*}

\begin{deluxetable}{cccccc}
	\tabletypesize{\scriptsize}
	\tablecaption{The CPU-scalability of the endian conversion alogorithms on Machine A\label{tab-endian-conversion-algorithm-scalabilityA}}
	\tablewidth{0pt}
	\tablehead{
		\colhead{Number of Threads} & \colhead{Bit Shift (msec)} & \colhead{\texttt{BSWAP} (msec)} & \colhead{SSE2 (msec)} & \colhead{SSSE3 (msec)} & \colhead{Byte Shuffle (msec)}
	}
	\startdata
		1 & $410 \pm 2$ & $411 \pm 3$ & $404 \pm 2$ & $372.4 \pm 0.2$ & $3191.8 \pm 0.7$ \\
		2 & $366 \pm 3$ & $366 \pm 3$ & $367 \pm 2$ & $360.4 \pm 0.5$ & $1605 \pm 2$ \\
		3 & $362 \pm 3$ & $362 \pm 3$ & $362 \pm 3$ & $354.9 \pm 0.3$ & $1082 \pm 6$ \\
		4 & $365 \pm 4$ & $365 \pm 4$ & $365 \pm 4$ & $358.8 \pm 0.5$ & $860 \pm 20$ \\
	\enddata
	\tablecomments{The CPU-scalability of 423,414,686 ($=$29,566$\times$14,321)
	\texttt{double}-type element endian conversion with various algorithms
	on Machine A.
	Different from Table~\ref{tab-endian-conversion-algorithm}, 16-byte memory
	alignment is adopted.
	}
\end{deluxetable}

\begin{deluxetable}{cccccc}
	\tabletypesize{\scriptsize}
	\tablecaption{The CPU-scalability of the endian conversion alogorithms on Machine B\label{tab-endian-conversion-algorithm-scalabilityB}}
	\tablewidth{0pt}
	\tablehead{
		\colhead{Number of Threads} & \colhead{Bit Shift (msec)} & \colhead{\texttt{BSWAP} (msec)} & \colhead{SSE2 (msec)} & \colhead{SSSE3 (msec)} & \colhead{Byte Shuffle (msec)}
	}
	\startdata
		1 & $597.8 \pm 0.7$ & $605.1 \pm 0.7$ & $579.3 \pm 0.9$ & $597 \pm 2$ & $8056.2 \pm 0.2$ \\
		2 & $484 \pm 1$ & $499 \pm 1$ & $484 \pm 1$ & $495 \pm 1$ & $4046 \pm 3$ \\
		3 & $429.9 \pm 0.8$ & $445 \pm 3$ & $432 \pm 1$ & $440 \pm 1$ & $2699 \pm 4$ \\
		4 & $413 \pm 3$ & $412 \pm 2$ & $413 \pm 3$ & $419 \pm 7$ & $2031 \pm 8$ \\
		5 & $418 \pm 3$ & $417 \pm 1$ & $422 \pm 2$ & $424 \pm 2$ & $1661 \pm 5$ \\
		6 & $412 \pm 1$ & $411.5 \pm 0.8$ & $413 \pm 1$ & $416 \pm 1$ & $1388 \pm 4$ \\
		7 & $412.3 \pm 0.9$ & $412.2 \pm 0.5$ & $411.9 \pm 0.6$ & $414.1 \pm 0.5$ & $1192 \pm 4$ \\
		8 & $416.1 \pm 0.6$ & $415.1 \pm 0.7$ & $415.5 \pm 0.4$ & $414.5 \pm 0.4$ & $1043 \pm 2$ \\
	\enddata
	\tablecomments{All conditions are same as Table~\ref{tab-endian-conversion-algorithm-scalabilityA}.}
\end{deluxetable}

\subsubsection{Multi-Thread \label{sec-endian-multithread}}

I also examined the CPU-scalability of these algorithms.
I adopted \texttt{pthread} for parallelization, and simply divided the array
containing the test data into equal-size segments so that the total number of
the segments were equal to the number of threads.
Then I assigned each thread with each segment.

Figure~\ref{fig-endian-algorithm} represents the results.
I also list the observed values for detailed comparison of the algorithms
in Table~\ref{tab-endian-conversion-algorithm-scalabilityA}
(for Machine A) and Table~\ref{tab-endian-conversion-algorithm-scalabilityB}
(for Machine B).
Except for byte shuffle algorithm, I observed $\simeq 10\%$
performance gain for Machine A, and $\simeq 40\%$ up to four threads
for Machine B with the four algorithms.

It seems strange that the memory bandwidth of Machine B is sufficient
for the test data size but the four algorithms show performance
cutoff at four threads.
I performed detailed hardware benchmark utilizing
LMbench\footnote{\texttt{http://www.bitmover.com/lmbench/}},
and found that context switching time and the latency of L2 cache memory
normalized in CPU cycles of Machine B are 2.4 times and 4.6 times,
respectively, larger than those of Machine A.
Hence I conclude that there are some hardware bottlenecks in Machine B,
which cause the plateau in Figure~\ref{fig-endian-algorithm}.

The behaviors of the four algorithms with respect to the number of
threads are very similar, and I adopt bit shift algorithm in the next
section because of its compiler portability and identicalness to \texttt{BSWAP}
(see Appendix~\ref{sec-bitshift-bswap}).

\section{Endian Conversion Timing}

A modern CPU has multiple arithmetic logic units (ALUs) and instruction
pipelines to boost the operating rates of ALUs.
As seen in the previous section, the hardware limitation lies just
below the endian conversion time of single thread
(Figure~\ref{fig-endian-algorithm}, Machine A), preventing
the CPU scalability.
This may lead to many holes (or ``no operation'' instructions)
in the pipelines and reduce the performance.
If this is the case, shuffling instructions in source codes can
produce improvement.

To verify this assumption, I disabled the endian conversion functionality
in CFITSIO;
I changed the \texttt{BYTESWAPPED} macros for i386 and amd64
architectures from \texttt{TRUE} into \texttt{FALSE} in \texttt{fitsio2.h},
and commented out the codes which CFITSIO perform runtime check to
verify whether the machine endian definition by the above macro is
consistent with the execution environment in \texttt{cfileio.c},
and I rebuilt the library.
The patches for those files are shown in Appendix~\ref{sec-cfitsio-patch}.

I compare the following two methods;
\begin{enumerate}
	\item It loads the full test image (Figure~\ref{fig-test-image}) from
		  tmpfs (see \S\ref{sec-configuration}) onto an array, then it
		  converts the endian by the parallelized bit shift algorithm
		  (described in \S\ref{sec-endian-multithread}), and it sums up
		  all the elements.
	\item It loads the full test image onto an array, and it sums up all
		  the elements with converting the endian one after another by
		  the bit shift algorithm.
\end{enumerate}
From here, I refer to the former as ``on ahead endian conversion
method'', and to the latter as ``just-in-time endian conversion method''.

On ahead endian conversion method can be written as follows:
\begin{verbatim}
{
  double *v;   // an array to store
               // a FITS image
  size_t len;  // the length of
               // the array v

  // load a byte sequence from a FITS
  // file into v here...
  
  // endian conversion
  for (size_t i = 0; i < len; ++i) {
    uint64_t *p = (uint64_t *)&v[i];
    uint64_t a = bit_shift(*p);
    double *q = (double *)&a;
    v[i] = *q;
  }

  // process v here...
}
\end{verbatim}
and just-in-time endian conversion method can be written as follows:
\begin{verbatim}
{
  double *v;   // an array to store
               // a FITS image
  size_t len;  // the length of
               // the array v

  // load a byte sequence from a FITS
  // file into v here...

  // image processing...
  {
    // something...
    
    // one needs to refer the value
    // of v[i] here
    
    // endian conversion
    uint64_t *p = (uint64_t *)&v[i];
    uint64_t a = bit_shift(*p);
    double *q = (double *)&a;
    double x = *q;
    // use x instead of v[i] below

    // something...
  }
}
\end{verbatim}
where \texttt{bit\_shift()} is the endian conversion
function defined in \S\ref{sec-definition-bit-shift}.

In this section, I adopt summing up all the elements in the test image
as an example of image processing.

\subsection{Single Thread}

I implemented both methods in single thread and performed
benchmark.
The codes of on ahead conversion method are following:
\begin{verbatim}
{
  // endian conversion
  for (size_t i = 0; i < len; ++i) {
    uint64_t *p = (uint64_t *)&v[i];
    uint64_t a = bit_shift(*p);
    double *q = (double *)&a;
    v[i] = *q;
  }

  // summation
  double sum = 0.0;
  for (size_t i = 0; i < len; ++i) {
    sum += v[i];
  }
}
\end{verbatim}
and those of just-in-time endian conversion method are following:
\begin{verbatim}
{
  double sum = 0.0;
  for (size_t i = 0; i < len; ++i) {
    // endian conversion
    uint64_t *p = (uint64_t *)&v[i];
    uint64_t a = bit_shift(*p);
    double *q = (double *)&a;

    // summation
    sum += *q;
  }
}
\end{verbatim}
Note that the former codes are identical to those with original CFITSIO.

The results are summarized in Table~\ref{tab-fitsprocess-single}.
I obtained slightly faster ($\simeq 5\%$) total processing time of
$2.22 \pm 0.04$ sec and $3.62 \pm 0.04$ sec for Machine A and B,
respectively, with on ahead
endian conversion method, while that with original CFITSIO is
$2.38 \pm 0.04$ and $3.79 \pm 0.03$ for Machine A and B, respectively.

On the other hand, I obtained significantly faster time of
$1.85 \pm 0.04$ and $3.05 \pm 0.03$ for Machine A and B, respectively,
which corresponds to $\simeq 25\%$ performance gain, with just-in-time endian
conversion method.

\subsection{Multi-Thread}

I made both methods multithreaded by utilizing
OpenMP\footnote{\url{http://openmp.org/wp/}} APIs for its simple
implementation.
The codes of just-in-time endian conversion method, for example, are below:
\begin{verbatim}
{
  double sum = 0.0;
  #pragma omp parallel for reduction (+:sum)\\
schedule (auto)
  for (size_t i = 0; i < len; ++i) {
    // endian conversion
    uint64_t *p = (uint64_t *)&v[i];
    uint64_t a = bit_shift(*p);
    double *q = (double *)&a;

    // summation
    sum += *q;
  }
}
\end{verbatim}

On the other hand, I could not find the best parameters in OpenMP APIs
for the endian conversion routine in on ahead conversion method, hence
I applied OpenMP only to the summation routine, and adopted
the \texttt{pthread}-based parallelization described in
\S\ref{sec-endian-multithread} for the endian conversion routine in on
ahead conversion method;
the number of the threads for OpenMP was set to that for the endian
conversion.

\begin{figure*}
	\begin{center}
		\includegraphics[keepaspectratio, width=0.75\hsize, clip]{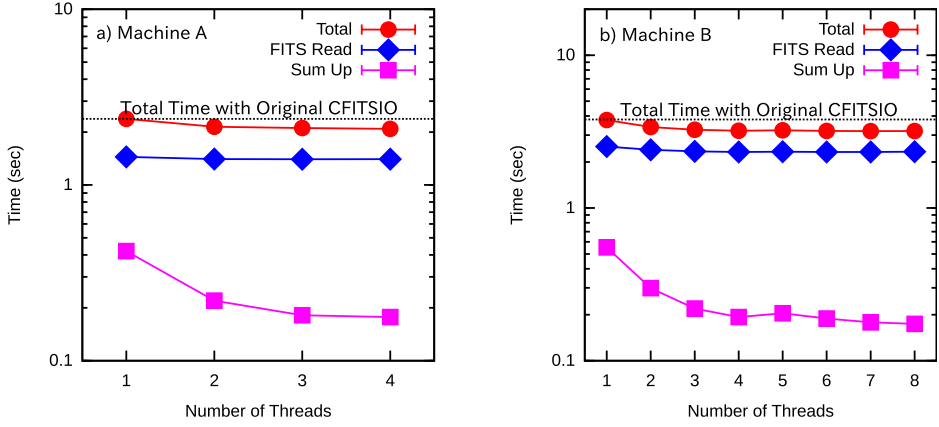}
	\end{center}
	\caption{The CPU-scalability of the FITS reading and processing time by applying
	the bit shift parallel algorithm to CFITSIO.
	I also applied a simple OpenMP parallelization to the summation routine.
	\label{fig-fits-processing}}
\end{figure*}

\begin{figure*}
	\begin{center}
		\includegraphics[keepaspectratio, width=0.75\hsize, clip]{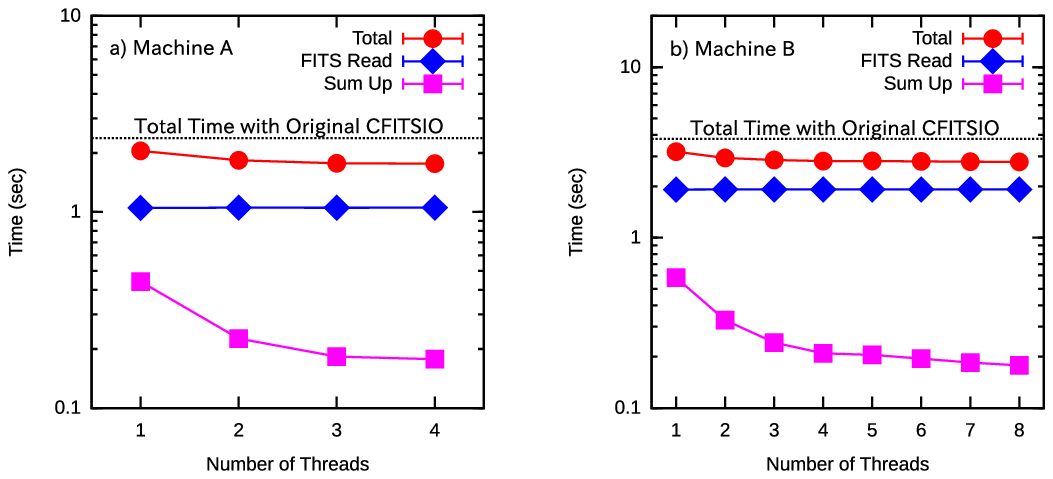}
	\end{center}
	\caption{The CPU-scalability of the just-in-time endian conversion algorithm.
	Endian conversions are performed in the summation routine, which is applied
	a simple OpenMP parallelization.
	\label{fig-fits-processing-lazy}}
\end{figure*}

The results obtained with these programs are summarized in
Table~\ref{tab-fitsprocess-multithread-a} (for Machine A),
Table~\ref{tab-fitsprocess-multithread-b} (for Machine B),
Figure~\ref{fig-fits-processing} (for on ahead endian conversion method),
and Figure~\ref{fig-fits-processing-lazy} (for just-in-time endian conversion method).
Note that the endian conversion time of on ahead endian conversion method
is included in the FITS reading time.
The total time to perform the same things with the original CFITSIO in
single thread is superimposed on these figures as a dotted line:
$2.38 \pm 0.04$ seconds for Machine A, and $3.79 \pm 0.03$ seconds for
Machine B.

For the on ahead endian conversion method, the total time slightly scales
the number of threads and gets faster than original CFITSIO, while the
file reading time (including endian conversion time) seems to be little
scalable.
The scalability of the total time mostly owes that of the summation
routine, and the parallelization of the endian conversion has little
impact due to the hardware limit seen in \S\ref{sec-endian-multithread}.

On the other hand, for the just-in-time endian conversion method, the total time
is interestingly smaller than that of original CFITSIO even for single
thread.
The summation routine seems to be scalable almost in the full range,
while the total time scales up to four threads.

\begin{deluxetable}{ccccc}
	\tabletypesize{\scriptsize}
	\rotate
	\tablecaption{The data processing times with two different method in single thread\label{tab-fitsprocess-single}}
	\tablewidth{0pt}
	\tablehead{
		\colhead{Method} & \colhead{Machine} & \colhead{FITS Read Time (sec)} & \colhead{Sum Up Time (sec)} & \colhead{Total Time (sec)}
	}
	\startdata
		 & Machine A & $1.006 \pm 0.003$ & $0.4198 \pm 0.0002$ & $2.22 \pm 0.04$ \\
		\raisebox{0.5em}{On Ahead Endian Conversion} & Machine B & $1.859 \pm 0.007$ & $0.557 \pm 0.001$ & $3.62 \pm 0.04$ \\ \hline
		 & Machine A & $1.010 \pm 0.003$ & $0.44003 \pm 0.00008$ & $1.85 \pm 0.04$ \\
		\raisebox{0.5em}{Lasy Endian Conversion} & Machine B & $1.874 \pm 0.007$ & $0.621 \pm 0.002$ & $3.06 \pm 0.03$ \\
	\enddata
	\tablecomments{The total time with original CFITSIO is $2.38 \pm 0.04$ and
	$3.79 \pm 0.03$ for Machine A and B, respectively.}
\end{deluxetable}

\begin{deluxetable}{ccccc}
	\tabletypesize{\scriptsize}
	\rotate
	\tablecaption{The CPU-scalability of on ahead and just-in-time endian conversion methods on Machine A\label{tab-fitsprocess-multithread-a}}
	\tablewidth{0pt}
	\tablehead{
		\colhead{Method} & \colhead{Number of Threads} & \colhead{FITS Read Time (sec)} & \colhead{Sum Up Time (sec)} & \colhead{Total Time (sec)}
	}
	\startdata
		 & 1 & $1.445 \pm 0.003$ & $0.4205 \pm 0.0001$ & $2.38 \pm 0.04$ \\
		 & 2 & $1.404 \pm 0.003$ & $0.220 \pm 0.004$ & $2.1 \pm 0.1$ \\
		\raisebox{0.5em}{On Ahead Endian Conversion Method} & 3 & $1.401 \pm 0.003$ & $0.181 \pm 0.002$ & $2.11 \pm 0.09$ \\
		 & 4 & $1.403 \pm 0.003$ & $0.177 \pm 0.002$ & $2.09 \pm 0.03$ \\ \hline
		 & 1 & $1.048 \pm 0.003$ & $0.4416 \pm 0.0009$ & $2.0 \pm 0.1$ \\
		 & 2 & $1.051 \pm 0.003$ & $0.2262 \pm 0.0005$ & $1.8 \pm 0.1$ \\
		\raisebox{0.5em}{Just-in-Time Endian Conversion Method} & 3 & $1.051 \pm 0.003$ & $0.183 \pm 0.001$ & $1.77 \pm 0.02$ \\
		 & 4 &  $1.052 \pm 0.003$ & $0.177 \pm 0.002$ & $1.76 \pm 0.02$ \\
	\enddata
	\tablecomments{The endian conversion time is included in the FITS reading time for on ahead endian conversion method.}
\end{deluxetable}

\begin{deluxetable}{ccccc}
	\tabletypesize{\scriptsize}
	\rotate
	\tablecaption{The CPU-scalability of on ahead and just-in-time endian conversion methods on Machine B\label{tab-fitsprocess-multithread-b}}
	\tablewidth{0pt}
	\tablehead{
		\colhead{Method} & \colhead{Number of Threads} & \colhead{FITS Read Time (sec)} & \colhead{Sum Up Time (sec)} & \colhead{Total Time (sec)}
	}
	\startdata
		 & 1 & $2.526 \pm 0.005$ & $0.5522 \pm 0.0006$ & $3.76 \pm 0.03$ \\
		 & 2 & $2.403 \pm 0.005$ & $0.299 \pm 0.003$ & $3.39 \pm 0.01$ \\
		 & 3 & $2.347 \pm 0.005$ & $0.220 \pm 0.002$ & $3.253 \pm 0.010$ \\
		 & 4 & $2.326 \pm 0.005$ & $0.193 \pm 0.003$ & $3.203 \pm 0.008$ \\
		\raisebox{0.5em}{On Ahead Endian Conversion Method} & 5 & $2.334 \pm 0.006$ & $0.205 \pm 0.003$ & $3.224 \pm 0.010$ \\
		 & 6 & $2.326 \pm 0.005$ & $0.189 \pm 0.003$ & $3.20 \pm 0.01$ \\
		 & 7 & $2.325 \pm 0.006$ & $0.178 \pm 0.002$ & $3.19 \pm 0.01$ \\
		 & 8 & $2.334 \pm 0.005$ & $0.174 \pm 0.003$ & $3.19 \pm 0.01$ \\ \hline
		 & 1 & $1.914 \pm 0.003$ & $0.582 \pm 0.002$ & $3.19 \pm 0.05$ \\
		 & 2 & $1.917 \pm 0.003$ & $0.328 \pm 0.002$ & $2.934 \pm 0.009$ \\
		 & 3 & $1.916 \pm 0.004$ & $0.242 \pm 0.002$ & $2.853 \pm 0.008$ \\
		 & 4 & $1.916 \pm 0.004$ & $0.205 \pm 0.002$ & $2.812 \pm 0.009$ \\
		\raisebox{0.5em}{Just-in-Time Endian Conversion Method} & 5 & $1.918 \pm 0.003$ & $0.205 \pm 0.001$ & $2.816 \pm 0.009$ \\
		 & 6 & $1.918 \pm 0.003$ & $0.1948 \pm 0.0009$ & $2.801 \pm 0.010$ \\
		 & 7 & $1.917 \pm 0.004$ & $0.185 \pm 0.001$ & $2.791 \pm 0.010$ \\
		 & 8 & $1.917 \pm 0.003$ & $0.178 \pm 0.003$ & $2.781 \pm 0.010$ \\
	\enddata
	\tablecomments{The endian conversion time is included in the FITS reading time for on ahead endian conversion method.}
\end{deluxetable}

\section{Discussion}

\subsection{Performance Analysis of the Simple Summation Codes}

There is a well-known equation to estimate the increase by parallelization,
Amdahl's law \citep{Amdahl1967}:
\begin{equation}
	T_{\rm parallel} = \left\{ \left( 1 - P \right) + \frac{P}{N} + \alpha \right\} T_{\rm single}, \label{eq-amdahl}
\end{equation}
where $T_{\rm single}$ and $T_{\rm parallel}$ represent processing time in
single thread and mult-thread cases, respectively, $P$ is the ratio of codes
which parallelization methods are applied to\footnote{Hardware bottlenecks
are included in the $1 - P$ term.}, $N$ is the number of threads,
and $\alpha$ is the overhead caused by parallelization.

\begin{figure*}
	\begin{center}
		\includegraphics[keepaspectratio, width=0.75\hsize, clip]{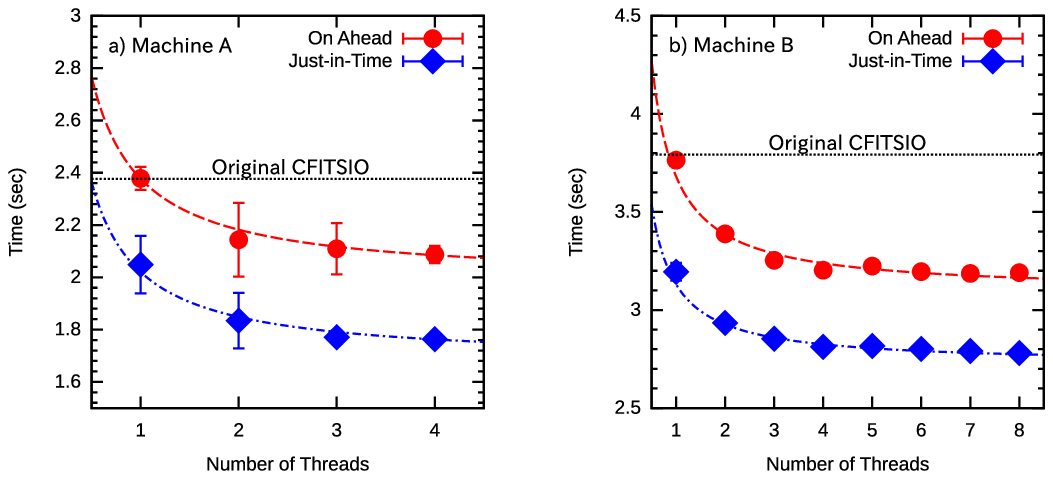}
	\end{center}
	\caption{The fitting results of on ahead and just-in-time endian conversion method
	with respect to total processing time with Amdahl's law.
	The red dashed and blue dash dotted lines correspond to the law for on ahead and
	just-in-time endian conversion method, respectively.
	\label{fig-amdahl}}
\end{figure*}

To quantify the performance increase of on ahead endian conversion method
and just-in-time endian conversion method, I performed model fitting to the total
time of both methods with Eq.(\ref{eq-amdahl}).
I found that $\alpha \sim O\left( 10^{-30} \right)$ while the fitting,
thus I fixed $\alpha$ at 0.
The results are summarized in Table~\ref{tab-endian-conversion-amdahl}
and Figure~\ref{fig-amdahl}.
The increasing rates of performance compared to original CFITSIO
($T_{\rm single} = 2.38 \pm 0.04$ for Machine A and $T_{\rm single} =
3.79 \pm 0.03$ for Machine B) are also listed in the table.

The figure shows that the above results are explained well
by Amdahl's law, and that the on ahead endian conversion method
for single thread has almost the same performance as original CFITSIO.
In fact, these two agree with each other in $\lesssim 5\%$ errors
according to the table.
The table also suggests that multi-threading boosts this method up
about 20\%.
Considering the parallelization rate $P \simeq 16\%$, one cannot
expect further speed up by multi-threading in $N \gtrsim 4$.
This suggests that the bottlenecks of other hardwares disrupt
order in the instruction pipelines and leads to the decrease of
operating ratio of ALUs.

On the other hand, the just-in-time endian conversion method is 20\% faster
than both of original CFITSIO and the single thread version of on
ahead one, surprisingly.
This seems as if the endian conversion process disappeared.
In the parallelized case, the just-in-time conversion method is 40\% faster
than the others in single thread.
However, the performance increase by multi-threading can be expected
only in $N \lesssim 4$ since the parallelization rate $P \simeq 16\%$,
due to the hardware bottlenecks mentioned above.

\begin{figure*}
	\begin{center}
		\includegraphics[keepaspectratio, width=0.75\hsize, clip]{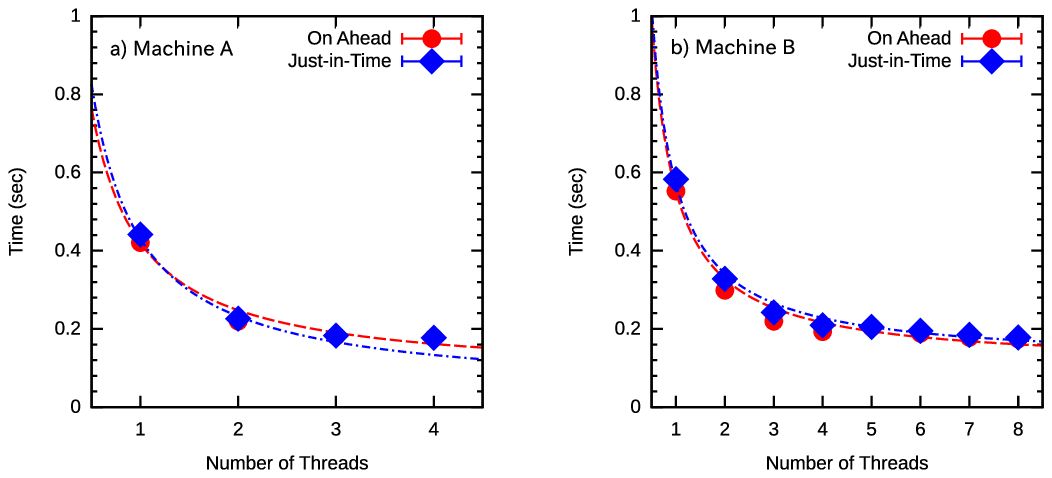}
	\end{center}
	\caption{The fitting results of on ahead and just-in-time endian conversion method
	with respect to summation time with Amdahl's law.
	The red dashed and blue dash dotted lines correspond to the law for on ahead and
	just-in-time endian conversion method, respectively.
	\label{fig-amdahl-sumup}}
\end{figure*}

For further investigation, I fitted the summation time of these methods
with Eq.(\ref{eq-amdahl}) to investigate the impact of the endian
conversion codes in the summation routine on performance; there are
endian conversion codes in the summation routine in case of just-in-time endian
conversion method, but not in case of on ahead endian conversion method.
The results are summarized in Table~\ref{tab-amdahl-sumup} and
Figure~\ref{fig-amdahl-sumup}.
I found that the parallelization rate $P \approx 85\%$ in both cases,
and that the ratio of $T_{\rm single}$ of just-in-time endian conversion
method against that of on ahead one
$r = T_{\rm single} \left( {\rm Just-in-Time} \right) / T_{\rm single} \left( {\rm On\ Ahead} \right)$
was equal to $r = 1.02 \pm 0.05$ for Machine A and $r = 1.03 \pm 0.04$
for Machine B.
There is no overhead of endian conversion in the summation routine,
since the shift of $r$ from unity is not significant statistically.

Thus I conclude that endian conversion is so simple operation for
a modern CPU that the bottlenecks of other hardwares disrupt order
in the instruction pipelines; to prevent the disruption, the endian
conversion should be done just before a value is referred.

\begin{deluxetable}{ccccccc}
	\tabletypesize{\scriptsize}
	\rotate
	\tablecaption{The fitting results of the total processing time\label{tab-endian-conversion-amdahl}}
	\tablehead{
		\colhead{} & \colhead{} & \colhead{} & \colhead{} & \colhead{} & \multicolumn{2}{c}{Increase Rate of Performance} \\
		\colhead{\raisebox{0.5em}{Method}} & \colhead{\raisebox{0.5em}{Machine}} & \colhead{\raisebox{0.5em}{$T_{\rm single}$ (sec)}} & \colhead{\raisebox{0.5em}{$P$}} & \colhead{\raisebox{0.5em}{$\chi^{2}$ (d.o.f.$^{\rm a}$)}} & \colhead{Single Thread} & \colhead{Multi-Thread}
	}
	\startdata
		 & Machine A & $2.377 \pm 0.009$ & $0.164 \pm 0.006$ & 0.09 (2) & $1.00 \pm 0.02$ & $1.20 \pm 0.02$ \\
		\raisebox{0.5em}{On Ahead Endian Conversion} & Machine B & $3.68 \pm 0.05$ & $0.16 \pm 0.02$ & 49$^{\rm b}$ (6) & $1.03 \pm 0.02$ & $1.22 \pm 0.03$ \\ \hline
		 & Machine A & $2.02 \pm 0.05$ & $0.17 \pm 0.03$ & 0.5 (2) & $1.18 \pm 0.04$ & $1.42 \pm 0.07$ \\
		\raisebox{0.5em}{Just-in-Time Endian Conversion} & Machine B & $3.13 \pm 0.02$ & $0.130 \pm 0.009$ & 8.5 (6) & $1.21 \pm 0.01$ & $1.39 \pm 0.02$ \\
	\enddata
	\tablecomments{The fitting results of the total processing time with respect to
	two different endian conversion methods with Amdahl's law and their increase
	rate of performance compared with original CFITSIO
	($T_{\rm single} = 2.38 \pm 0.04$ for Machine A and $T_{\rm single}
	= 3.79 \pm 0.03$ for Machine B).
	The errors are 1-$\sigma$ confidence limits for a single parameter.}
	\tablenotetext{a}{Degrees of freedom}
	\tablenotetext{b}{The large $\chi^{2}$ value is caused by very small errors of observed values,
    which agree with the model well (see Figure~\ref{fig-amdahl}).}
\end{deluxetable}

\begin{deluxetable}{ccccc}
	\tabletypesize{\scriptsize}
	\rotate
	\tablecaption{The fitting results of the time to sum up all elements\label{tab-amdahl-sumup}}
	\tablehead{
		\colhead{Method} & \colhead{Machine} & \colhead{$T_{\rm single}$ (msec)} & \colhead{$P$} & \colhead{$\chi^{2}$ (d.o.f.)}
	}
	\startdata
		 & Machine A & $420 \pm 1$ & $0.82 \pm 0.04$ & 127.9 (2) \\
		\raisebox{0.5em}{On Ahead Endian Conversion} & Machine B & $551 \pm 6$ & $0.81 \pm 0.02$ & 491.1 (6) \\ \hline
		 & Machine A & $430 \pm 20$ & $0.92 \pm 0.06$ & 913.4 (2) \\
		\raisebox{0.5em}{Just-in-Time Endian Conversion} & Machine B & $570 \pm 20$ & $0.80 \pm 0.02$ & 312.7 (6) \\
	\enddata
	\tablecomments{$T_{\rm single} \left( {\rm Just-in-Time} \right) / T_{\rm single} \left( {\rm On\ Ahead} \right) = 1.02 \pm 0.05$ (for Machine A), $1.03 \pm 0.04$ (for Machine B).}
\end{deluxetable}

\begin{figure}
	\includegraphics[keepaspectratio, width=0.9\hsize,clip]{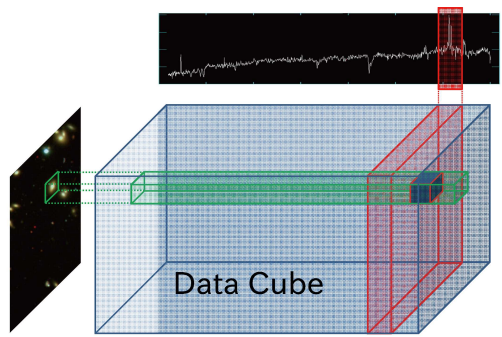}
	\caption{The schematic illustration of a data cube of ALMA
	Science Verification Data.\label{fig-alma-data-cube}}
\end{figure}

\begin{deluxetable}{ccc}
	\tabletypesize{\scriptsize}
	\tablecaption{Image Extraction Time\label{tab-almawebql-image}}
	\tablehead{
		\colhead{File Size (MB)} & \colhead{On Ahead Endian Conversion Method
	} & \colhead{Just-in-Time Endian Conversion Method}
	}
	\startdata
		$21.4$ & $0.02953 \pm 0.00006$ & $0.0361 \pm 0.0004$ \\
		$55.0$ & $0.75214 \pm 0.00003$ & $0.0910 \pm 0.0001$ \\
		$66.9$ & $0.10007 \pm 0.00006$ & $0.10295 \pm 0.00008$ \\
		$78.4$ & $0.13136 \pm 0.00008$ & $0.12371 \pm 0.00008$ \\
		$88.7$ & $0.14435 \pm 0.00007$ & $0.1299 \pm 0.0001$ \\
		$106.9$ & $0.1835 \pm 0.0001$ & $0.1606 \pm 0.0002$ \\
		$150.3$ & $0.2333 \pm 0.0001$ & $0.21006 \pm 0.00008$
	\enddata
	\tablecomments{The time to extract an image from an ALMA data cube on Machine A in single thread.}
\end{deluxetable}

\begin{deluxetable}{ccc}
	\tabletypesize{\scriptsize}
	\tablecaption{Spectrum Extraction Time\label{tab-almawebql-spectrum}}
	\tablehead{
		\colhead{File Size (MB)} & \colhead{On Ahead Endian Conversion Method
	} & \colhead{Just-in-Time Endian Conversion Method}
	}
	\startdata
		$21.4$ & $0.02743 \pm 0.00002$ & $0.0344 \pm 0.0004$ \\
		$55.0$ & $0.06970 \pm 0.00002$ & $0.07527 \pm 0.00008$ \\
		$66.9$ & $0.08512 \pm 0.00002$ & $0.08583 \pm 0.00010$ \\
		$78.4$ & $0.09861 \pm 0.00003$ & $0.0909 \pm 0.0001$ \\
		$88.7$ & $0.11271 \pm 0.00003$ & $0.10539 \pm 0.00008$ \\
		$106.9$ & $0.13394 \pm 0.00004$ & $0.1153 \pm 0.0003$ \\
		$150.3$ & $0.18769 \pm 0.00005$ & $0.15503 \pm 0.00006$
	\enddata
	\tablecomments{The time to extract a spectrum from an ALMA data cube on Machine A in single thread.}
\end{deluxetable}

\begin{figure*}
	\begin{center}
		\includegraphics[keepaspectratio, width=0.75\hsize,clip]{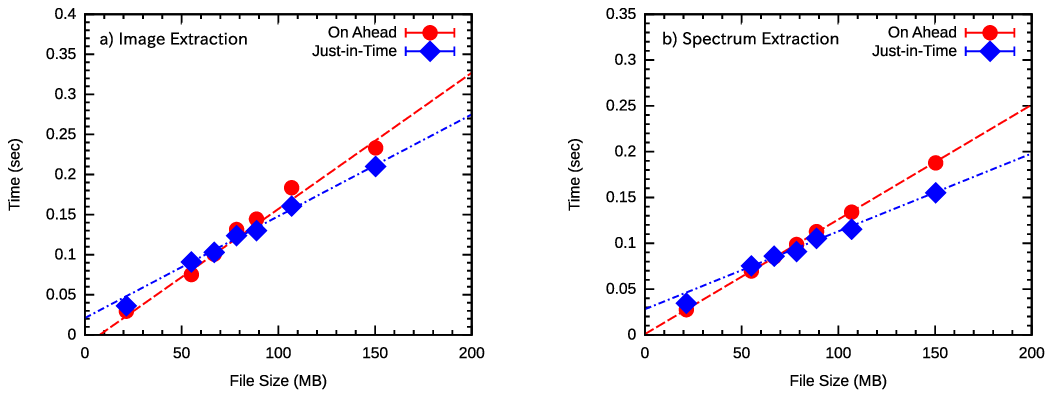}
	\end{center}
	\caption{The image (left) and spectrum (right) extraction time with respect to file size of ALMA data cube.
	The red dashed and blue dash dotted lines correspond to the best fit
	ones for on ahead and just-in-time endian conversion method,
	respectively.\label{fig-almawebql}}
\end{figure*}

\subsection{Application to ALMAWebQL}

From here, I only investigated the performance increase of summing
up all the elements in a large FITS file by just-in-time endian
conversion method.
In this subsection, I apply the method to ALMAWebQL, our interactive web
viewer for ALMA data cubes described in Paper I, to obtain more realistic
benchmark data.
For realistic and fair comparison, the SSE2 boosted endian conversion
codes in CFITSIO are enabled for on ahead endian conversion method,
while there is no SSE2 code in just-in-time endian conversion method.

ALMA data cubes not contain information of polarization currently,
and they are simple 3-dimensional FITS files (Figure~\ref{fig-alma-data-cube}).
For image extraction, one have to integrate the cube along the spectral
direction; for spectrum extraction, one convolute all spatial information.

I measured the time to complete these computations in single thread
with various size data on Machine A.
The results for image extraction are summarized in Table~\ref{tab-almawebql-image}, and those for spectrum extraction are
summarized in Table~\ref{tab-almawebql-spectrum}.
From these tables, I obtain
\begin{eqnarray}
T \left( {\rm On\ Ahead} \right) & = & \left( 1.7 \pm 0.1 \right) \times 10^{-3} \left( \frac{V}{\rm MB} \right) \nonumber \\
& & - \left( 0.013 \pm 0.007 \right) \  \sec,
\end{eqnarray}
\begin{eqnarray}
T \left( {\rm Just-in-Time} \right) & = &\left( 1.27 \pm 0.04 \right) \times 10^{-3} \left( \frac{V}{\rm MB} \right) \nonumber \\
& & + \left( 0.021 \pm 0.003 \right) \ \sec
\end{eqnarray}
for image extraction and
\begin{eqnarray}
T \left( {\rm On\ Ahead} \right) & = & \left( 1.252 \pm 0.007 \right) \times 10^{-3} \left( \frac{V}{\rm MB} \right) \nonumber \\
& & + \left( 0.0008 \pm 0.0004 \right) \ \sec,
\end{eqnarray}
\begin{eqnarray}
T \left( {\rm Just-in-Time} \right) & = &\left( 0.85 \pm 0.02 \right) \times 10^{-3} \left( \frac{V}{\rm MB} \right) \nonumber \\
& & + \left( 0.028 \pm 0.002 \right) \ \sec
\end{eqnarray}
for spectrum extraction, where $T \left( {\rm On\ Ahead} \right)$ and
$T \left( {\rm Just-in-Time} \right)$ represent the time with on ahead and
just-in-time endian conversion methods, respectively, and $V$ is file size
in the MB unit (Figure~\ref{fig-almawebql}).
Hence just-in-time endian conversion method in single thread is
$\gtrsim 20\%$ faster than on ahead conversion method boosted by SSE2
above $V \gtrsim 200 \ \rm{MB}$.
This demonstrates that just-in-time endian conversion method can be very
powerful when one performs convolution and stacking of very large images,
which are very common analysis techniques in optical band, obtained with
future large telescopes.

\subsection{Data Types}

In this paper, I only treated a double precision FITS file,
but one could expect almost the same results for float and
\texttt{LONG} data types, which correspond to \texttt{BITPIX} $=$ -32 and
32, respectively;
as demonstrated in Appendix~\ref{sec-bitshift-bswap}, \texttt{bit\_shift()}
function is compiled into \texttt{BSWAP} instruction.
The amd64 architecture can handle both of 32 bit and 64 bit
operation codes and their operands seamlessly.
On the other hand, for byte and short data type, there may be little
advantage of just-in-time endian conversion method since \texttt{BSWAP}
instruction cannot take any 16 bit values as its operand,
and up-casting into 32 bit integer always occurs in arithmetic
operations in both cases.

\section{Summary}

The FITS format was originally developed to exchange digital
astronomical datasets from a computer to another, but the progress
of computation power and software technology enables one to process
FITS files through web browsers.
In addition, data size has been inflating year by year, and it will
exceed $\sim$\ TB in the year ahead.
To handle such big FITS file with web applications, the endian
conversion time from the FITS native to the machine one cannot be
negligible, and a solution for this problem is required.

In this paper, I compared the features of four typical endian conversion
algorithms under multi-thread environment, and found the bit shift one
was suitable for parallelization.
Then I examined the best timing for endian conversion under multi-thread
environment.
I found that one should postpone the endian conversion until a value
is really referred in a program, because endian conversion is so simple
for a modern CPU that the bottlenecks of other hardwares disrupt order
in the instruction pipelines, which leads to the decrease of operating
ratio of ALUs.
In fact, by applying this method to loading 3.4\ GB FITS file and sum
up all the elements, the performance increased 20\% for single thread
and 40\% for multi-thread compared to CFITSIO, which corresponded to
$\gtrsim 600$ milliseconds, and one can be aware of the speed-up.
No overhead of endian conversion was found on the summation routine;
hence one can sweep the endian conversion time out of his/her codes.
Note that parallelization of this method peaked out in four threads
in the experiment.

CPU vendors introduce various techniques, such as speculative
execution and branch prediction, to improve the efficiency of instruction
pipelines; an executed instruction code sequence is apart from a
programmed one.
In this context, modern CPUs partially break ``causality'', a programmed
instruction code sequence, and gain speed.
Just-in-time endian conversion method utilizes such boosting technology.
There is nothing new in the method, but it must be a small step to
handle astronomical big data generated by the next generation
telescopes.

\acknowledgments
I greatly appreciate Dr.~Chisato Yamauchi, who is my colleague and the author
of SLLIB/SFITSIO\footnote{SFITSIO is a light weight FITS library for C/C++, providing
modern APIs.},
for rewarding discussions.

\appendix

\section{Tmpfs and Ramdisk \label{sec-tmpfs}}

Both tmpfs and ramdisk are a data space allocated on memory.
One has to specify the size in advance for ramdisk, while
one does not set the size for tmpfs in advance necessarily
since it is under control of virtual memory manager and shares
swap space.

When an application requests the operating system for memory
blocks and when there does not remain sufficient physical memory
space, the memory manager firstly swap out the files on tmpfs.
Tmpfs is ideal space to put temporal files which one requires
very fast access to.

\section{Another Implementation of Byte Shuffle Algorithm \label{sec-another-byteshuffle}}

One can also implements byte shuffle algorithm as follows:
\begin{verbatim}
uint64_t byte_shuffle2(uint64_t a)
{
  unsigned char *p = (unsigned char *)&a;
  uint64_t b;
  unsigned char *q = (unsigned char *)&b;

  q[0] = p[7];
  q[1] = p[6];
  q[2] = p[5];
  q[3] = p[4];
  q[4] = p[3];
  q[5] = p[2];
  q[6] = p[1];
  q[7] = p[0];

  return b;
}
\end{verbatim}
The number of assignments of the codes ($= 8$) is less than that shown
in the main part of this paper ($= 12$), and one would expect further
performance improvement.

I disassembled both two codes compiled with the \texttt{-O2} option,
and obtained followings:
\begin{verbatim}
0000000000000000 <byte_shuffle>:
   0:	49 89 fa             	mov    %rdi,%r10
   3:	49 89 f8             	mov    %rdi,%r8
   6:	89 fe                	mov    %edi,%esi
   8:	48 89 f9             	mov    %rdi,%rcx
   b:	89 fa                	mov    %edi,%edx
   d:	48 89 f8             	mov    %rdi,%rax
  10:	40 88 7c 24 ff       	mov    %dil,-0x1(%rsp)
  15:	48 c1 e8 20          	shr    $0x20,%rax
  19:	49 c1 ea 38          	shr    $0x38,%r10
  1d:	49 c1 e8 30          	shr    $0x30,%r8
  21:	66 c1 ee 08          	shr    $0x8,%si
  25:	48 c1 e9 28          	shr    $0x28,%rcx
  29:	c1 ea 10             	shr    $0x10,%edx
  2c:	c1 ef 18             	shr    $0x18,%edi
  2f:	44 88 54 24 f8       	mov    %r10b,-0x8(%rsp)
  34:	40 88 74 24 fe       	mov    %sil,-0x2(%rsp)
  39:	44 88 44 24 f9       	mov    %r8b,-0x7(%rsp)
  3e:	88 54 24 fd          	mov    %dl,-0x3(%rsp)
  42:	88 4c 24 fa          	mov    %cl,-0x6(%rsp)
  46:	40 88 7c 24 fc       	mov    %dil,-0x4(%rsp)
  4b:	88 44 24 fb          	mov    %al,-0x5(%rsp)
  4f:	48 8b 44 24 f8       	mov    -0x8(%rsp),%rax
  54:	c3                   	retq   
\end{verbatim}
, and
\begin{verbatim}
0000000000000000 <byte_shuffle2>:
   0:	49 89 fa             	mov    %rdi,%r10
   3:	49 89 f9             	mov    %rdi,%r9
   6:	49 89 f8             	mov    %rdi,%r8
   9:	48 89 fe             	mov    %rdi,%rsi
   c:	89 f9                	mov    %edi,%ecx
   e:	89 fa                	mov    %edi,%edx
  10:	89 f8                	mov    %edi,%eax
  12:	49 c1 ea 38          	shr    $0x38,%r10
  16:	49 c1 e9 30          	shr    $0x30,%r9
  1a:	66 c1 e8 08          	shr    $0x8,%ax
  1e:	49 c1 e8 28          	shr    $0x28,%r8
  22:	48 c1 ee 20          	shr    $0x20,%rsi
  26:	c1 e9 18             	shr    $0x18,%ecx
  29:	c1 ea 10             	shr    $0x10,%edx
  2c:	44 88 54 24 f8       	mov    %r10b,-0x8(%rsp)
  31:	44 88 4c 24 f9       	mov    %r9b,-0x7(%rsp)
  36:	44 88 44 24 fa       	mov    %r8b,-0x6(%rsp)
  3b:	40 88 74 24 fb       	mov    %sil,-0x5(%rsp)
  40:	88 4c 24 fc          	mov    %cl,-0x4(%rsp)
  44:	88 54 24 fd          	mov    %dl,-0x3(%rsp)
  48:	88 44 24 fe          	mov    %al,-0x2(%rsp)
  4c:	40 88 7c 24 ff       	mov    %dil,-0x1(%rsp)
  51:	48 8b 44 24 f8       	mov    -0x8(%rsp),%rax
  56:	c3                   	retq   
\end{verbatim}
, that is, there are less assignments in \texttt{byte\_shuffle2()} ($= 8$)
than \texttt{byte\_shuffle()} ($= 12$), however, the former binary codes
are longer than the latter ones.
Hence one cannot expect more performance gain with the codes.

\section{Bit Shift Algorithm and \texttt{BSWAP} Instruction \label{sec-bitshift-bswap}}

The bit shift endian conversion codes is actually identical to \texttt{BSWAP}
instruction when compiled with the optimization option of \texttt{-O2}.
The disassembled codes obtained with \texttt{objdump -d} command
are below:
\begin{verbatim}
0000000000000000 <bit_shift>:
   0:   48 89 f8                mov    %rdi,%rax
   3:   48 0f c8                bswap  %rax
   6:   c3                      retq   
\end{verbatim}

\begin{deluxetable}{cccccc}
	\tabletypesize{\scriptsize}
	\tablecaption{The endian conversion time in case that memory alignment is random\label{tab-endian-conversion-algorithm-ramdom}}
	\tablehead{
		\colhead{Machine} & \colhead{Bit Shift (msec)} & \colhead{\texttt{BSWAP} (msec)} & \colhead{SSE2 (msec)} & \colhead{SSSE3 (msec)} & \colhead{Byte Shuffle (msec)}
	}
	\startdata
		Machine A & $413 \pm 2$ & $416 \pm 3$ & $416 \pm 5$ & $377 \pm 2$ & $3220 \pm 10$ \\
		Machine B & $624 \pm 7$ & $641 \pm 9$ & $602 \pm 9$ & $590 \pm 3$ & $8320 \pm 90$ \\
	\enddata
	\tablecomments{The endian conversion time of 423,414,686 ($=$29,566$\times$14,321)
	\texttt{double}-type elements with various algorithms in case that
	memory alignment is random.
	}
\end{deluxetable}

\section{Endian Conversion Algorithms and Memory Alignment}

It is ensured that the leading memory address (alignment) of an array
is always in multiplies of 16 (16-byte alignment) in amd64 architecture.
However, if one would like to read a file in multi-thread, he/she has
to make the copy of the file image on memory.
In such case, the alignment is not always 16-byte.
Thus I performed the benchmark described in \S\ref{sec-endian-benchmark}
(single thread case) but made alignment of the arraya random number.

For the benchmark, I modified \texttt{\_mm\_load\_si128()} and
\texttt{\_mm\_store\_si128()} in the SSE2 codes into
\texttt{\_mm\_loadu\_si128()} and \texttt{\_mm\_storeu\_si128()},
respectively, to make the codes operable.
The results are summarized in
Table~\ref{tab-endian-conversion-algorithm-ramdom}.

The trend found in \S\ref{sec-endian-single-thread} is roughly
true in this case though the all algorithms are slightly slower
(within a few \%) than 16-byte alignment case.
Hence one does not have to get nervous about memory alignment.

\section{The Patches for CFITSIO \label{sec-cfitsio-patch}}

\subsection{\texttt{fitsio2.h}}

\begin{verbatim}
*** fitsio2.h.org	2013-03-08 14:19:49.560538980 +0900
--- fitsio2.h	2013-01-15 14:43:10.000000000 +0900
***************
*** 96,102 ****
  
  #elif defined(__ia64__)  || defined(__x86_64__)
                    /*  Intel itanium 64-bit PC, or AMD opteron 64-bit PC */
! #define BYTESWAPPED TRUE
  #define LONGSIZE 64   
  
  #elif defined(_SX)             /* Nec SuperUx */
--- 96,103 ----
  
  #elif defined(__ia64__)  || defined(__x86_64__)
                    /*  Intel itanium 64-bit PC, or AMD opteron 64-bit PC */
! /* #define BYTESWAPPED TRUE */
! #define BYTESWAPPED FALSE
  #define LONGSIZE 64   
  
  #elif defined(_SX)             /* Nec SuperUx */
***************
*** 169,175 ****
  
  /*  generic 32-bit IBM PC */
  #define MACHINE IBMPC
! #define BYTESWAPPED TRUE
  
  #elif defined(__arm__)
  
--- 170,178 ----
  
  /*  generic 32-bit IBM PC */
  #define MACHINE IBMPC
! /* #define BYTESWAPPED TRUE */
! #define BYTESWAPPED FALSE
! 
  
  #elif defined(__arm__)
  
\end{verbatim}

\subsection{\texttt{cfileio.c}}

\begin{verbatim}
*** cfileio.c.org	2013-03-08 14:20:09.052539296 +0900
--- cfileio.c	2013-01-16 19:57:46.000000000 +0900
***************
*** 3763,3769 ****
      }
  
      /*   test for correct byteswapping.   */
! 
      u.ival = 1;
      if  ((BYTESWAPPED && u.cval[0] != 1) ||
           (BYTESWAPPED == FALSE && u.cval[1] != 1) )
--- 3763,3769 ----
      }
  
      /*   test for correct byteswapping.   */
! /*
      u.ival = 1;
      if  ((BYTESWAPPED && u.cval[0] != 1) ||
           (BYTESWAPPED == FALSE && u.cval[1] != 1) )
***************
*** 3776,3782 ****
        FFUNLOCK;
        return(1);
      }
!     
      
      /*  test that LONGLONG is an 8 byte integer */
      
--- 3776,3782 ----
        FFUNLOCK;
        return(1);
      }
! */    
      
      /*  test that LONGLONG is an 8 byte integer */
      

\end{verbatim}

%% figures
\clearpage

%% tables
\clearpage

\end{document}